\documentclass[aps,prl,
floatfix,
twocolumn,
longbibliography,
showpacs,
superscriptaddress
]{revtex4-1}
\usepackage{amsmath,amssymb}
\usepackage{graphicx}
\usepackage[ansinew]{inputenc}
\usepackage{hyperref,color}

\newcommand\abinitio{{\it ab initio~}} 

\newcommand\siesta{\textsc{Siesta}}
\newcommand\tsiesta{\textsc{TranSiesta}}

\begin{document}

\author{Jing-Tao\ L\"u}
\affiliation{School of Physics and Wuhan National High Magnetic Field Center, Huazhong University of Science and Technology, 430074 Wuhan, China}
\author{Susanne\ Leitherer}
\affiliation{Department of Physics, Technical University of Denmark, DK-2800 Kongens Lyngby, Denmark}
\author{Nick\ R.\ Papior}
\affiliation{Department of Applied Mathematics and Computer Science, Technical University of Denmark, DK-2800 Kongens Lyngby, Denmark}
\author{Mads\ Brandbyge}
\affiliation{Department of Physics, Technical University of Denmark, DK-2800 Kongens Lyngby, Denmark}

\title{{Ab initio} current-induced molecular dynamics}

\begin{abstract}
We extend the \abinitio molecular dynamics (AIMD) method based on density functional theory to the nonequilibrium situation where an electronic current is present in the electronic system. The dynamics is treated using the semi-classical generalized Langevin equation.
We demonstrate how the full anharmonic description of the inter-atomic forces is important in order to understand the current-induced heating and the energy distribution both in frequency and in real space.
\end{abstract}


\maketitle


Energy dissipation from an electronic current into the motion of atoms is an important and challenging scientific and technological problem\cite{Seideman2016,Pop2010}.
Especially, conductors consisting of a few atoms can be driven into a highly nonequilibrium vibrational state by a current\cite{Galperin2007,Todorov1998}. 
The local excitation of the atomic vibrations are hard to address in experiments\cite{Ward2011,Ioffe2008,Huang2006,Schirm2013,Smit2004,Agrait2002} and theoretically\cite{Todorov2001,Dundas2009,Frederiksen04,Lu2010}. 
So far the formation of atomic "hot-spots" has been demonstrated using first principles calculations at the level of density functional theory (DFT) using the harmonic approximation thus neglecting inter-mode coupling\cite{Lu2015}. It was also shown how current-induced momentum transfer ("electron wind" forces) may displace the "hot-spots" up or down-stream w.r.t. the current. This behavior has been observed in recent experiments\cite{Lee2013,Tsutsui2018}. 
In a simplified picture one may consider the atomic motion in the conductor to have a higher effective temperature compared to surrounding electrodes\cite{Todorov1998,Tsutsui2005}. However, the fact that individual vibrational modes couple differently to the electronic current and phonon reservoirs makes this picture invalid in general. Moreover, the anharmonic inter-mode coupling distributes the energy among these and makes the problem challenging to treat beyond model calculations.

Here we present a computational method which allows for \abinitio molecular dynamics (AIMD) simulation of atomic scale conductors in the presence of electronic current. It can address the effects of the current on the vibrational excitation (Joule heating), the current-induced "wind" forces, the coupling to electrode phonon reservoirs\cite{Engelund2010}, and the anharmonic couplings between the modes of the conductor\cite{Wang07}. We demonstrate the method on two "bench-mark" atomic scale conductors, an atomic gold chain and a benzene molecule, and analyze the role of the individual effects for the dissipation. 

\emph{Langevin equation.--}
We consider an atomic scale conductor consisting of a central  bottleneck junction and two electrodes, "left" and "right", respectively. In the presence of an electrical current, electrons are scattered near the junction and deposit energy to the surrounding atoms. We are interested in the atomic dynamics driven by the electrical current across the junction. To this end, we define the "system" as the coordinates of the nuclei in the junction. All remaining degrees of freedom (DOF) are treated as reservoirs/baths. We consider two kinds of reservoirs. The first is the electronic one, including all the electron DOF in the conductor and electrodes. Importantly, in the presence of electrical current, the electron reservoir is in a nonequilibrium state. This non-thermal reservoir serves as the energy source, depositing energy to the atoms. The second kind is the phonon reservoirs of the left and right electrode parts.

Employing the Feynman-Vernon influence functional approach, a semi-classical generalized Langevin equation (SCGLE) can be derived to describe the dynamics of the nonequilibrium system,\cite{Lu2010,Bode2011,Lu2012,Kantorovich2018,Chen2019,Lu2019}
\begin{align}
 \bm{\ddot Q} (t) = \bm{F}( \bm{Q}(t)) +\int_{-\infty}^t \bm{\Pi^r} (t-t') \bm{Q}(t') dt' + \bm{f}(t).
 \label{eq:lang}
\end{align}
Here, $\bm{Q}(t)$ is the mass-normalized displacement of system nuclei away from their equilibrium positions, i.e., $Q_i(t) = \sqrt{m_i} (R_i(t)-R^0_i)$, with $i$ representing the nuclear and its Cartesian index, $m_i$ the nuclear mass, $R_i(t)$ and $R^0_i$ the nuclear positions at time $t$ and at equilibrium, respectively. 
We split the forces acting on the system into different terms.
The force, $\bm{F}(\bm{Q}(t))$, is obtained from equilibrium DFT, while the corrections due to coupling to reservoirs and the effects of nonequilibrium are reflected in the last two terms at the right hand side of Eq.~(\ref{eq:lang}). We treat this latter coupling perturbatively to linear order in $Q$. Thus we limit ourselves to quasi-stationary situations where the system is stable on the time-scale much longer than the vibrations and the excursions from equilibrium are still small.
We neglect here the constant (zero order in $Q$) current-induced force which can be trivially included and focus on the excitation effects due to the other terms.  

The time non-local term describes the back-action of reservoirs on the system due to finite system displacement $\bm{Q}$, while $\bm{f}$ is the fluctuating/stochastic force with zero mean. It can be characterized by the correlation function,
\begin{align}
\langle \bm{f}(t)\bm{f}^T(t')\rangle =\hbar \bm{\hat\Pi}(t-t')\,,
\end{align}
which includes thermal and quantum fluctuations. 
Both $\bm{\hat\Pi}$ and $\bm{\Pi^r}$ are given by a sum of contributions, i.e., $\bm{\Pi^r} = \bm{\Pi}^r_e +\bm{\Pi}^r_{L}+\bm{\Pi}^r_R$, with $e$, $L$ and $R$ representing the electron, the left and the right phonon reservoirs, respectively. 

In nonequilibrium $\Pi^r$ and $\hat{\Pi}$ can be obtained using nonequilibrium Green's functions\cite{Lu2012}. Importantly, they depend on the presence of current: The bias-dependent part of $\hat{\Pi}_e$ generates Joule heating, while $\Pi^r_e$ contains current-induced forces. In nonequilibrium the fluctuation-dissipation relation linking $\Pi^r$ and $\hat{\Pi}$ becomes invalid, and the individual magnitudes of $\Pi^r$ and $\hat{\Pi}$ matters. This is opposed to equilibrium AIMD where one may more freely choose how to describe the coupling to reservoirs which in the end thermalize the system. 

Previously Eq.~(\ref{eq:lang}) has been used for \abinitio calculations in the harmonic approximation\cite{Lu2010,Lu2012,Lu2015}. Here, we go beyond this limitation and perform AIMD simulations by numerically solving Eq.~(\ref{eq:lang}), with all parameters obtained from the same DFT setup\footnote{We use SIESTA\cite{Soler2002}, TranSIESTA\cite{Brandbyge2002} and Inelastica \cite{Frederiksen07} tool set to get the electronic, vibrational spectrum and their coupling. For the MD, we use the i-PI software\cite{ipi} (cf. SM)}.
In the following we consider two prototype atomic conductors and illustrate the importance of anharmonic vibrational coupling on the energy redistribution both in frequency and in real space.

\begin{figure}[]
 \centering
       \includegraphics[width=0.49\textwidth]{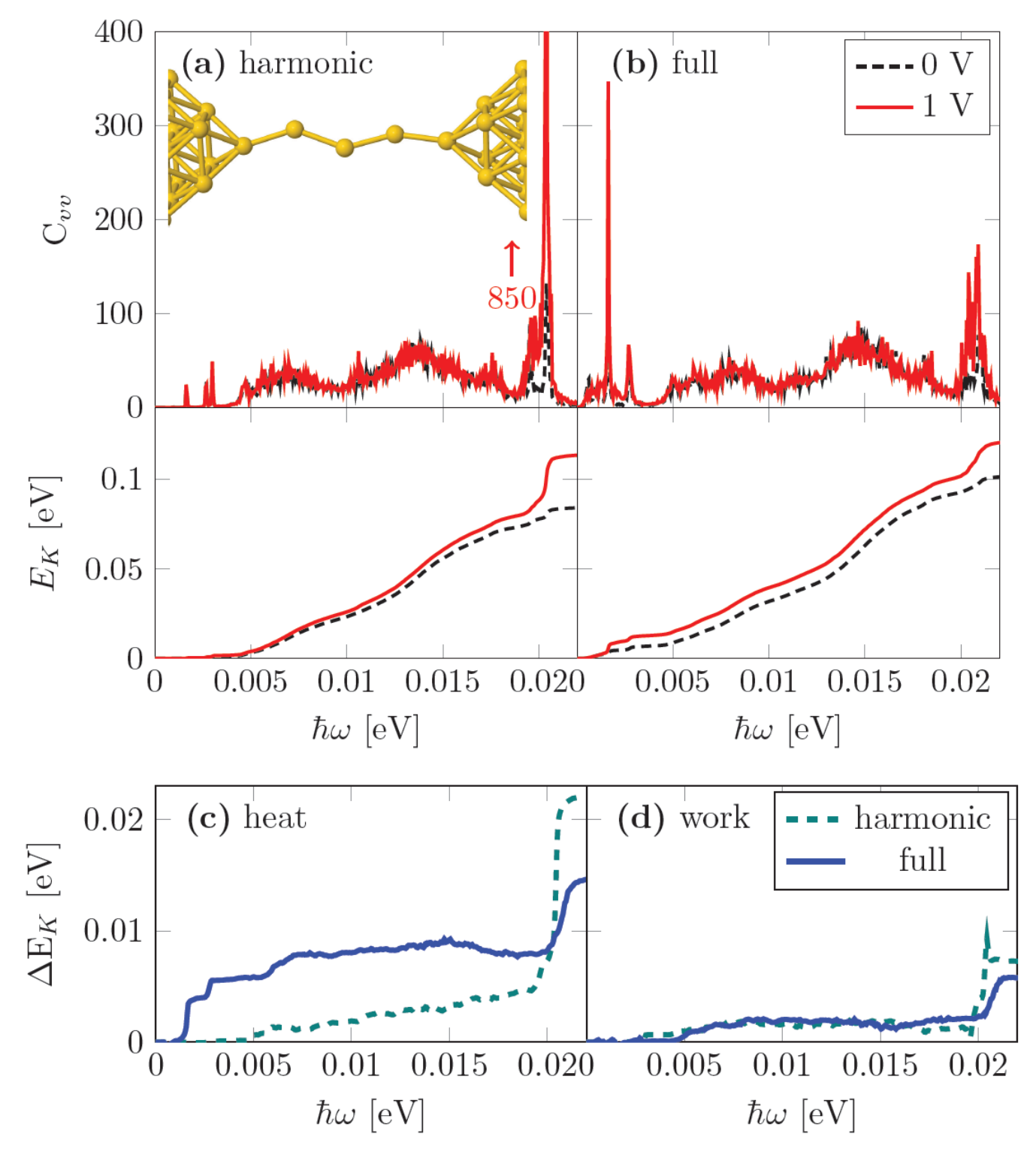}
      \caption{Power spectra ($C_{vv}$) and its cumulative integral ($E_K$) of the 5-atom Au chain connecting to Au(100) electrodes from MD calculations (inset). (a) Harmonic calculation using the dynamical matrix. The inset shows the structure of the chain. The Au(100) electrodes at 4K with a cross section of $3\time 3$ atoms are not shown.  (b) Full AIMD calculation. (c-d) The excess kinetic energy separated into contributions of heat and work. The work is contributed by the current-induced NC force, while the heat is from Joule heating. The coupling of low frequency modes $<$4 meV to the electrons was excluded in the calculation. We use a time step of $10$ fs in the MD simulation.} 
      \label{fig:heating}
\end{figure}
\emph{Au chain --}
The first example is a 5-atom gold chain between two Au(100) electrodes [inset of Fig.~\ref{fig:heating} (a)]. The temperature of the electrodes are 4K. 
In the harmonic approximation, Eq.~(\ref{eq:lang}) is linear and can be solved directly. We first check that the MD result in the harmonic approximation re-produces that from direct solution of Eq.~(\ref{eq:lang}) (Fig.~S1).
After validating the method, we proceed to study energy transfer from the nonequilibrium electronic bath to the system when the applied voltage bias is finite, $V\neq 0$. There are two energy transfer channels: The first is the stochastic Joule heating from the bias dependent fluctuating force correlation function $\hat{\Pi}_e$. The second is the work done to the system through the energy non-conserving\cite{Dundas2009} (NC) current-induced forces in $\Pi^r_e$. The energy deposited in the system can be further transferred into the left/right equilibrium phonon reservoirs. If the system remains stable it reaches a steady state when the energy into the system is balanced with the energy out to the phonon reservoirs.

\begin{figure}[b!]
\centering
\includegraphics[width=0.5\textwidth]{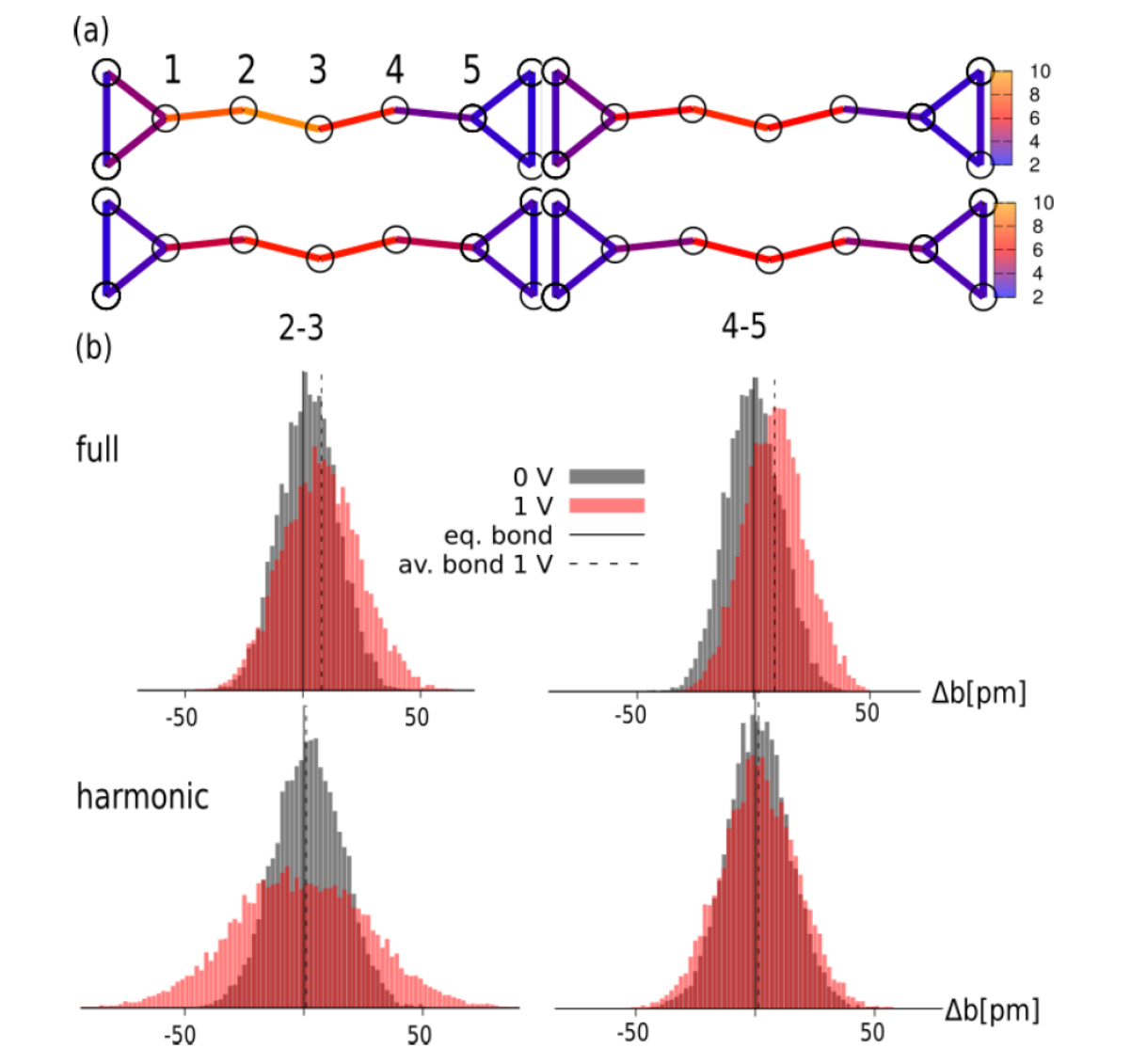} 
            \caption{ (a) Comparison of bond kinetic energy in meV from harmonic (left) and full AIMD (right) calculations. Shown is the bond energy at 1V with (top, asymmetric) and without (bottom, symmetric) NC force.
            (b) Length distribution of bond between atoms 2-3 and 4-5 calculated from the
             full AIMD and the harmonic approximation at 1 V. The distribution of bond 2-3 with high kinetic energy is strongly broadened at 1 V, with maximum $\Delta b=0.1$ \AA. 
            }
            \label{fig:au-map}
\end{figure}

After reaching a steady state, we use the power spectrum of the velocity-velocity correlation function $C_{vv}$ (Eq.~\ref{eq:cvv}) to characterize the frequency/energy-resolved contribution to the system kinetic energy. Integrating $C_{vv}$ yields the kinetic energy stored in the system for frequencies below $\omega$,
\begin{equation}
E_K(\omega) = \int_0^{\omega} \frac{d\omega'}{2\pi} {C}_{vv}(\omega')\,.
\label{eq:ek}
\end{equation}
The calculated results for $C_{vv}$ along with $E_K$ are shown in Fig.~\ref{fig:heating}(a) for the harmonic calculation using the dynamical matrix from DFT, and in Fig.~\ref{fig:heating}(b) for the full AIMD calculation. 
Applying a bias voltage results in an increase in vibrational kinetic energy, which can be divided into separate heat and work contributions, shown in  Fig.~\ref{fig:heating}(c) and (d), respectively. The work part comes from the contribution of NC current-induced forces, while the heat part is from Joule heating.

Comparing results at $V={\rm 0 V}$ (equilibrium) and $V={\rm 1 V}$ (nonequilibrium) from the harmonic approximation, we find that electrons transfer energy mainly to the high frequency modes at $\sim 0.02$eV. This is indicated by a big step at $\sim 0.02$eV at 1V [red line in lower panel of Fig.~\ref{fig:heating} (a)]. These modes are the alternating-bond-length (ABL) modes consistent with previous experimental and theoretical studies\cite{Agrait2002,Frederiksen04,Frederiksen07,Lu2015}. In addition to stronger coupling to electrons, the ABL frequencies are located above the bulk phonon band of Au. This hinders direct harmonic energy transfer to the phonon reservoirs. Inclusion of the anharmonic coupling leads to an energy redistribution from the high to the low frequency modes. This is seen from the strongly enhanced heating of the low frequency modes ($< 5$ meV) in the anharmonic result, which is absent in the harmonic case. The anharmonic coupling also reduces the excess kinetic energy in the system by opening an indirect channel of energy transfer from the ABL modes to the bulk electrode phonon reservoirs [Fig.~\ref{fig:heating} (c-d)].

We now analyze the MD results in real space to gain further understanding of the bond properties. 
In Fig.~\ref{fig:au-map}(a) the kinetic energy per bond [Eq.~\ref{eq:bond}] is shown for the harmonic and anharmonic cases. In each case, we compare the situations with and without work contribution from non-conservative (NC) current-induced forces.
The inclusion of NC current-induced forces leads to a left-right symmetry breaking in the real space energy distribution. Specifically, more energy is accumulated in bonds near the left electrode from where the current enters the junction. 
This asymmetric concentration of energy ('hot spot') was explained by the asymmetric excitation of left- and right-traveling phonon waves by NC current-induced forces\cite{Lu2015}. 

Due to this effect, some bonds in the chain heat up strongly. We have compared the statistics of bond fluctuations between atoms 2-3 and 4-5 in both the harmonic and full AIMD calculations [Fig.~\ref{fig:au-map} (b)]. The 2-3 bond shows larger broadening compared to the 4-5 bond in both cases. This is a direct result of asymmetric heating. In the AIMD calculation, the broadening is reduced compared to the harmonic result. This is due to the anharmonic energy transport channel to the phonon reservoirs, consistent with the analysis in Fig.~\ref{fig:heating}. Moreover, we observe elongation of both bonds in the AIMD result. This is another manifestation of anharmonicity\footnote{In order to clearly observe the effect we have excluded the change of the atomic structure and thus the bond length\cite{Brandbyge2003}.}.

\begin{figure}[]
 \centering
 \includegraphics[width=0.5\textwidth]{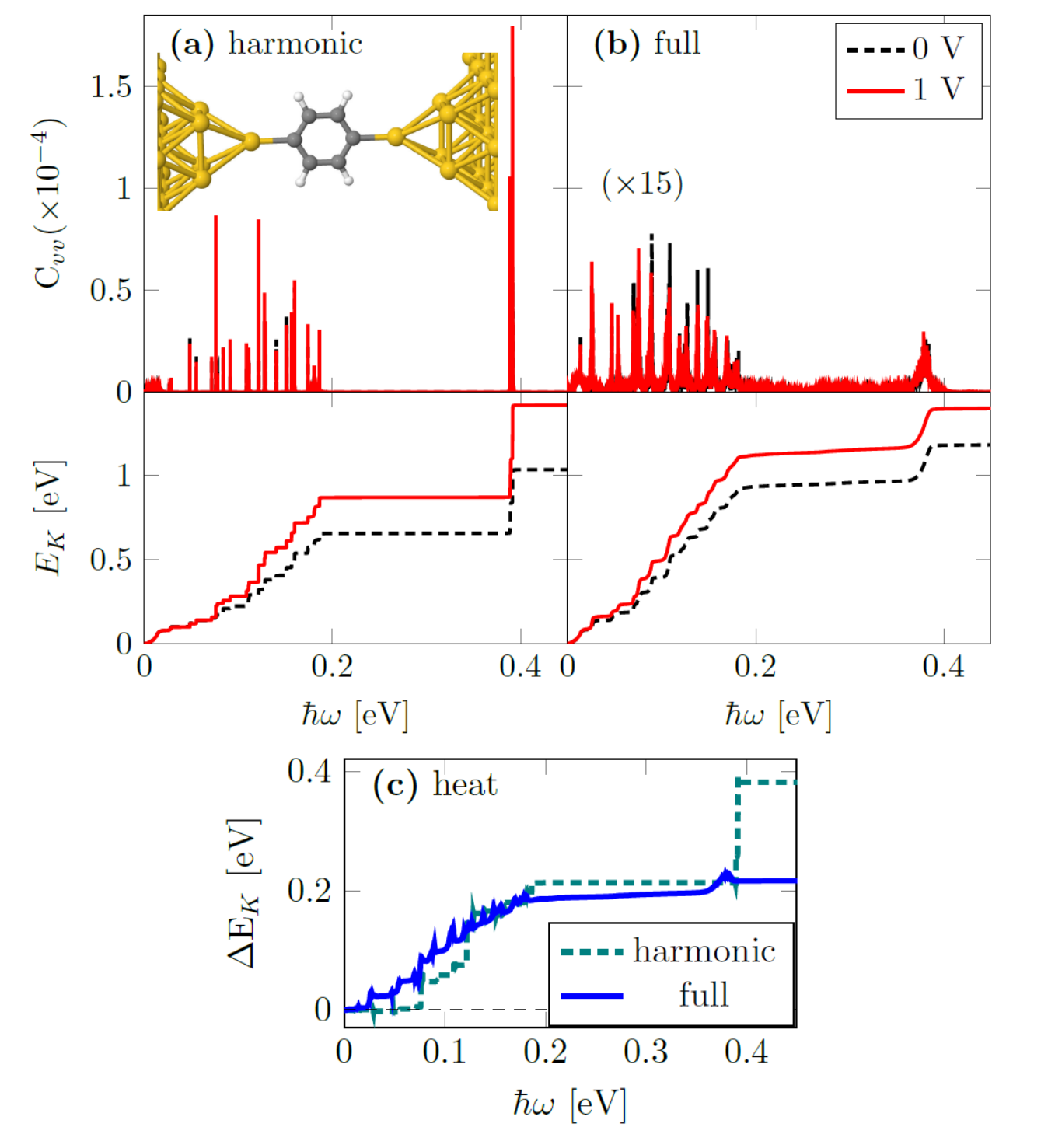}
\caption{Power spectra ($C_{vv}$) and its cumulative integral ($E_K$) of the Benzene molecule connecting to Au electrodes through Au-C bonds.  (a) Harmonic calculation using the dynamical matrix. The inset shows the junction geometry. The Au(100) electrodes with a cross section of $4\times 4$ atoms are not shown. (b) Anharmonic calculation. (c) Comparison of excess kinetic energy from harmonic and anharmonic calculations. The contribution of work from the NC force is negligible in this case. }
\label{fig:ben}
\end{figure}

\emph{Benzene molecular junction --}
The second example we consider is a Benzene molecular junction connecting to  Au(100) electrodes at 4K through direct Au-C bonds\cite{Cheng2011} [inset of Fig.~\ref{fig:ben} (a)]. Due to different atomic mass and bonding type, there is a large frequency mismatch between Benzene vibrational modes ($0-380$ meV) and the gold electrode phonons ($0-18$ meV). Only low frequency vibrational modes ($<18$ meV)  couple directly to electrode phonons. The remaining high frequency modes ($> 18$ meV) appear as sharp peaks in the power spectrum [Fig.~\ref{fig:ben} (a)]. Work done by current-induced forces in this junction is negligible due to low electrical conductance. The energy increase at $1$ V is dominated by Joule heating. It involves mainly vibrational modes larger than $75$ meV. Including the anharmonic vibrational coupling leads to large spectrum broadening [Fig.~\ref{fig:ben} (b)], and an excess heating much smaller than the harmonic case [Fig.~\ref{fig:ben} (c)]. Similar to the Au chain, we observe heating of the vibrational modes below $75$ meV. This
is a direct result of anharmonic vibrational coupling between different modes. 

%

\begin{figure}[h!]
	\centering
\includegraphics[width=0.5\textwidth]{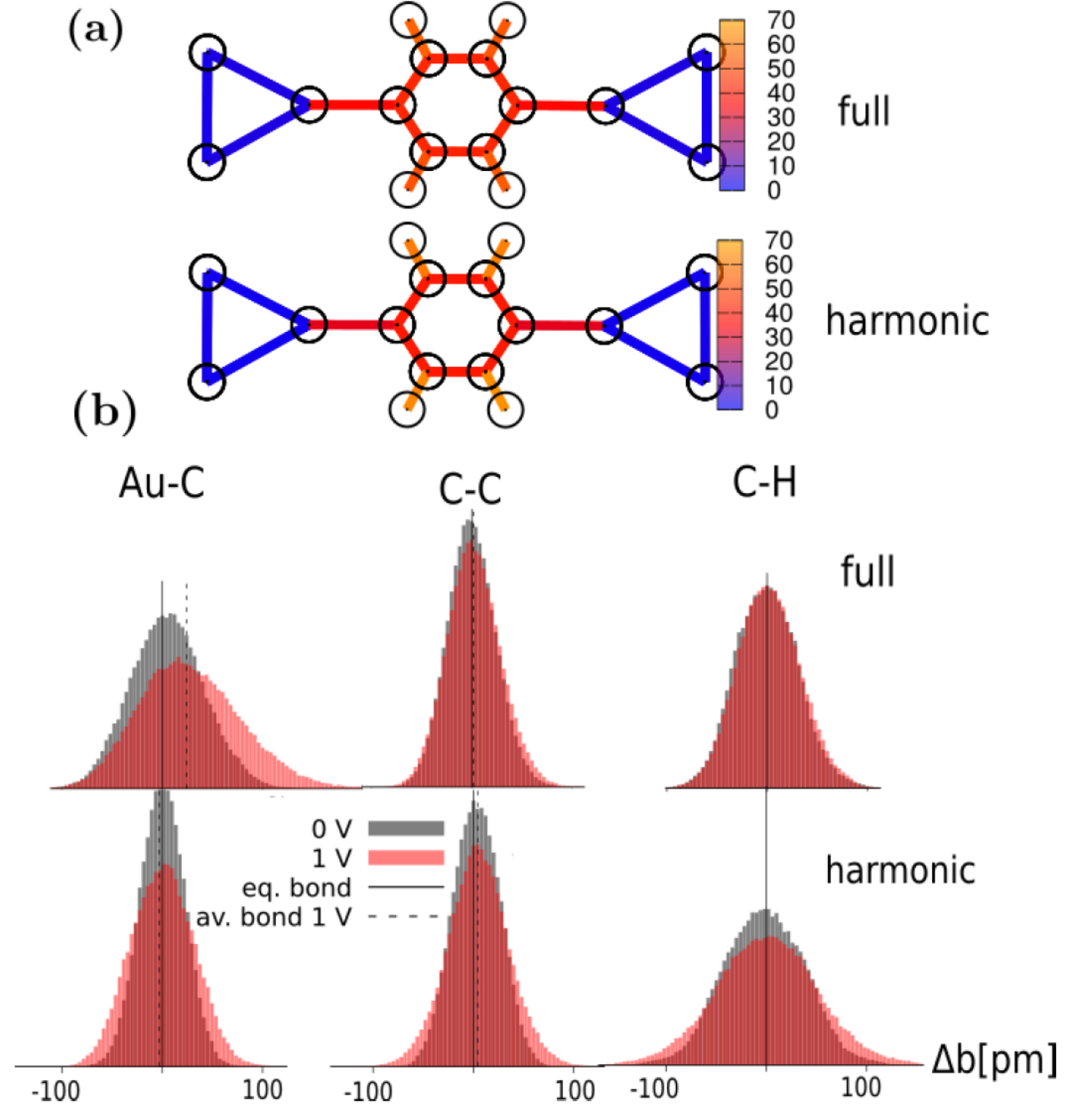}
	\caption{(a) Comparison of kinetic energy per bond in meV from the full AIMD and the harmonic calculation.
	(b) Length distribution of different bonds calculated from the two methods. In the full AIMD results, the Au-C bond exhibits the widest distribution with a maximum bond elongation of $\Delta b=0.2$ \AA~ and the largest shift of the mean bond length with bias. Interestingly, it also shows certain skewness with larger probability to longer length side.}
	\label{fig:ben-bonds}
\end{figure}

The real space distribution of kinetic energy per bond at 1 V  is shown in Fig.~\ref{fig:ben-bonds}(a). It is symmetric due to the symmetry of the junction and the negligible contribution of NC current-induced forces.
The bond length histograms for the different types of bonds in the junction are depicted in Fig.~\ref{fig:ben-bonds}(b) for harmonic (lower panel) and full AIMD (upper panel) calculations. Comparing the two cases, we find that, the excess energy of C-H and C-C bonds is effectively transferred away to the Au-C bond in the presence of anharmonic coupling. 
This is evidenced from the reduced broadening of C-C and C-H bonds and the enhanced broadening of Au-C bond in the full AIMD results. As a result, the average Au-C bond length increases by $\sim 0.2$~\AA. More interestingly, the Au-C bond distribution shows obvious skewness, thus the average length is larger than the most probable length.  This is an important signature of bond anharmonicity, and can only be probed by the full AIMD calculation.

In experiments an
effective temperature, $T_{\rm eff}$, has been obtained from contact life-times\cite{Tsutsui2008} or two-level conductance fluctuations\cite{Tsutsui2007} via an Arrhenius type rate expression.
The $T_{\rm eff}$ has been fitted to a $V$-dependence\cite{Todorov1998}, as $T_{\rm eff}=\gamma \sqrt{V}$ (zero ambient temperature). 
For both systems above we do not find the energy 
distribution among the modes is compatible with an overall effective temperature (cf. Fig.~ \ref{fig:teff}). 
On the other hand we find a maximum kinetic energy of interatomic Au-Au (Au-C) bonds corresponding to 80 K (320 K) which is comparable to experimental values\cite{Tsutsui2007,Tsutsui2008} of 75 K (450 K).

In summary, we have presented a method for performing \abinitio molecular dynamics of an atomic scale conductor in the presence of an electrical current. We illustrated this approach by studying two prototype molecular conductors and analyzing the subtle interplay between Joule heating, current-induced forces and anharmonic vibrational coupling among different modes. We found that anharmonic coupling among different modes has an important effect on the energy redistribution both in frequency and in real space. Our study opens the possibility of studying current-induced atomic switches, chemical reactions and bond ruptures\cite{Erp2018,solomon2017} beyond simple models.

We acknowledge funding support from National Natural Science Foundation of China (Grant No. 21873033), the National Key Research and Development Program of China (Grant No. 2017YFA0403501), the program for HUST academic frontier youth team and Villum Fonden (Grant No. 00013340). 
We thank the Shanghai supercomputer center for providing computing resources.

\bibliographystyle{apsrev4-1}
\bibliography{Heating,CurrentInducedForces,my}

\begin{thebibliography}{38}%
\makeatletter
\providecommand \@ifxundefined [1]{%
 \@ifx{#1\undefined}
}%
\providecommand \@ifnum [1]{%
 \ifnum #1\expandafter \@firstoftwo
 \else \expandafter \@secondoftwo
 \fi
}%
\providecommand \@ifx [1]{%
 \ifx #1\expandafter \@firstoftwo
 \else \expandafter \@secondoftwo
 \fi
}%
\providecommand \natexlab [1]{#1}%
\providecommand \enquote  [1]{``#1''}%
\providecommand \bibnamefont  [1]{#1}%
\providecommand \bibfnamefont [1]{#1}%
\providecommand \citenamefont [1]{#1}%
\providecommand \href@noop [0]{\@secondoftwo}%
\providecommand \href [0]{\begingroup \@sanitize@url \@href}%
\providecommand \@href[1]{\@@startlink{#1}\@@href}%
\providecommand \@@href[1]{\endgroup#1\@@endlink}%
\providecommand \@sanitize@url [0]{\catcode `\\12\catcode `\$12\catcode
  `\&12\catcode `\#12\catcode `\^12\catcode `\_12\catcode `\%12\relax}%
\providecommand \@@startlink[1]{}%
\providecommand \@@endlink[0]{}%
\providecommand \url  [0]{\begingroup\@sanitize@url \@url }%
\providecommand \@url [1]{\endgroup\@href {#1}{\urlprefix }}%
\providecommand \urlprefix  [0]{URL }%
\providecommand \Eprint [0]{\href }%
\providecommand \doibase [0]{http://dx.doi.org/}%
\providecommand \selectlanguage [0]{\@gobble}%
\providecommand \bibinfo  [0]{\@secondoftwo}%
\providecommand \bibfield  [0]{\@secondoftwo}%
\providecommand \translation [1]{[#1]}%
\providecommand \BibitemOpen [0]{}%
\providecommand \bibitemStop [0]{}%
\providecommand \bibitemNoStop [0]{.\EOS\space}%
\providecommand \EOS [0]{\spacefactor3000\relax}%
\providecommand \BibitemShut  [1]{\csname bibitem#1\endcsname}%
\let\auto@bib@innerbib\@empty
\bibitem [{\citenamefont {Seideman}(2016)}]{Seideman2016}%
  \BibitemOpen
  \bibfield  {author} {\bibinfo {author} {\bibfnamefont {T.}~\bibnamefont
  {Seideman}},\ }\href {https://books.google.co.jp/books?id=rZGNaUNrBRAC}
  {\emph {\bibinfo {title} {Current-Driven Phenomena in Nanoelectronics}}}\
  (\bibinfo  {publisher} {Jenny Stanford Publishing},\ \bibinfo {year}
  {2016})\BibitemShut {NoStop}%
\bibitem [{\citenamefont {Pop}(2010)}]{Pop2010}%
  \BibitemOpen
  \bibfield  {author} {\bibinfo {author} {\bibfnamefont {E.}~\bibnamefont
  {Pop}},\ }\bibfield  {booktitle} {\emph {\bibinfo {booktitle} {Nano
  Research}},\ }\href {\doibase 10.1007/s12274-010-1019-z} {\ \textbf {\bibinfo
  {volume} {3}},\ \bibinfo {pages} {147} (\bibinfo {year} {2010})}\BibitemShut
  {NoStop}%
\bibitem [{\citenamefont {Galperin}\ \emph {et~al.}(2007)\citenamefont
  {Galperin}, \citenamefont {Ratner},\ and\ \citenamefont
  {Nitzan}}]{Galperin2007}%
  \BibitemOpen
  \bibfield  {author} {\bibinfo {author} {\bibfnamefont {M.}~\bibnamefont
  {Galperin}}, \bibinfo {author} {\bibfnamefont {M.~A.}\ \bibnamefont
  {Ratner}}, \ and\ \bibinfo {author} {\bibfnamefont {A.}~\bibnamefont
  {Nitzan}},\ }\href@noop {} {\bibfield  {journal} {\bibinfo  {journal}
  {Journal of Physics Condensed Matter}\ }\textbf {\bibinfo {volume} {19}},\
  \bibinfo {pages} {103201} (\bibinfo {year} {2007})}\BibitemShut {NoStop}%
\bibitem [{\citenamefont {Todorov}(1998)}]{Todorov1998}%
  \BibitemOpen
  \bibfield  {author} {\bibinfo {author} {\bibfnamefont {T.~N.}\ \bibnamefont
  {Todorov}},\ }\href {\doibase 10.1080/13642819808206398} {\bibfield
  {journal} {\bibinfo  {journal} {Phil. Mag. B}\ }\textbf {\bibinfo {volume}
  {77}},\ \bibinfo {pages} {965} (\bibinfo {year} {1998})}\BibitemShut
  {NoStop}%
\bibitem [{\citenamefont {Ward}\ \emph {et~al.}(2011)\citenamefont {Ward},
  \citenamefont {Corley}, \citenamefont {Tour},\ and\ \citenamefont
  {Natelson}}]{Ward2011}%
  \BibitemOpen
  \bibfield  {author} {\bibinfo {author} {\bibfnamefont {D.~R.}\ \bibnamefont
  {Ward}}, \bibinfo {author} {\bibfnamefont {D.~A.}\ \bibnamefont {Corley}},
  \bibinfo {author} {\bibfnamefont {J.~M.}\ \bibnamefont {Tour}}, \ and\
  \bibinfo {author} {\bibfnamefont {D.}~\bibnamefont {Natelson}},\ }\href
  {\doibase 10.1038/nnano.2010.240} {\bibfield  {journal} {\bibinfo  {journal}
  {Nat. Nanotechnol.}\ }\textbf {\bibinfo {volume} {6}},\ \bibinfo {pages} {33}
  (\bibinfo {year} {2011})}\BibitemShut {NoStop}%
\bibitem [{\citenamefont {Ioffe}\ \emph {et~al.}(2008)\citenamefont {Ioffe},
  \citenamefont {Shamai}, \citenamefont {Ophir}, \citenamefont {Noy},
  \citenamefont {Yutsis}, \citenamefont {Kfir}, \citenamefont {Cheshnovsky},\
  and\ \citenamefont {Selzer}}]{Ioffe2008}%
  \BibitemOpen
  \bibfield  {author} {\bibinfo {author} {\bibfnamefont {Z.}~\bibnamefont
  {Ioffe}}, \bibinfo {author} {\bibfnamefont {T.}~\bibnamefont {Shamai}},
  \bibinfo {author} {\bibfnamefont {A.}~\bibnamefont {Ophir}}, \bibinfo
  {author} {\bibfnamefont {G.}~\bibnamefont {Noy}}, \bibinfo {author}
  {\bibfnamefont {I.}~\bibnamefont {Yutsis}}, \bibinfo {author} {\bibfnamefont
  {K.}~\bibnamefont {Kfir}}, \bibinfo {author} {\bibfnamefont {O.}~\bibnamefont
  {Cheshnovsky}}, \ and\ \bibinfo {author} {\bibfnamefont {Y.}~\bibnamefont
  {Selzer}},\ }\href {\doibase 10.1038/nnano.2008.304} {\bibfield  {journal}
  {\bibinfo  {journal} {Nat. Nanotechnol.}\ }\textbf {\bibinfo {volume} {3}},\
  \bibinfo {pages} {727} (\bibinfo {year} {2008})}\BibitemShut {NoStop}%
\bibitem [{\citenamefont {Huang}\ \emph {et~al.}(2006)\citenamefont {Huang},
  \citenamefont {Xu}, \citenamefont {Chen}, \citenamefont {{Di Ventra}},\ and\
  \citenamefont {Tao}}]{Huang2006}%
  \BibitemOpen
  \bibfield  {author} {\bibinfo {author} {\bibfnamefont {Z.}~\bibnamefont
  {Huang}}, \bibinfo {author} {\bibfnamefont {B.}~\bibnamefont {Xu}}, \bibinfo
  {author} {\bibfnamefont {Y.}~\bibnamefont {Chen}}, \bibinfo {author}
  {\bibfnamefont {M.}~\bibnamefont {{Di Ventra}}}, \ and\ \bibinfo {author}
  {\bibfnamefont {N.}~\bibnamefont {Tao}},\ }\href {\doibase 10.1021/nl0608285}
  {\bibfield  {journal} {\bibinfo  {journal} {Nano Lett.}\ }\textbf {\bibinfo
  {volume} {6}},\ \bibinfo {pages} {1240} (\bibinfo {year} {2006})}\BibitemShut
  {NoStop}%
\bibitem [{\citenamefont {Schirm}\ \emph {et~al.}(2013)\citenamefont {Schirm},
  \citenamefont {Matt}, \citenamefont {Pauly}, \citenamefont {Cuevas},
  \citenamefont {Nielaba},\ and\ \citenamefont {Scheer}}]{Schirm2013}%
  \BibitemOpen
  \bibfield  {author} {\bibinfo {author} {\bibfnamefont {C.}~\bibnamefont
  {Schirm}}, \bibinfo {author} {\bibfnamefont {M.}~\bibnamefont {Matt}},
  \bibinfo {author} {\bibfnamefont {F.}~\bibnamefont {Pauly}}, \bibinfo
  {author} {\bibfnamefont {J.~C.}\ \bibnamefont {Cuevas}}, \bibinfo {author}
  {\bibfnamefont {P.}~\bibnamefont {Nielaba}}, \ and\ \bibinfo {author}
  {\bibfnamefont {E.}~\bibnamefont {Scheer}},\ }\href {\doibase
  10.1038/nnano.2013.170} {\bibfield  {journal} {\bibinfo  {journal} {Nat.
  Nanotechnol.}\ }\textbf {\bibinfo {volume} {8}},\ \bibinfo {pages} {645}
  (\bibinfo {year} {2013})}\BibitemShut {NoStop}%
\bibitem [{\citenamefont {Smit}\ \emph {et~al.}(2004)\citenamefont {Smit},
  \citenamefont {Untiedt},\ and\ \citenamefont {{Van Ruitenbeek}}}]{Smit2004}%
  \BibitemOpen
  \bibfield  {author} {\bibinfo {author} {\bibfnamefont {R.~H.}\ \bibnamefont
  {Smit}}, \bibinfo {author} {\bibfnamefont {C.}~\bibnamefont {Untiedt}}, \
  and\ \bibinfo {author} {\bibfnamefont {J.~M.}\ \bibnamefont {{Van
  Ruitenbeek}}},\ }\href@noop {} {\bibfield  {journal} {\bibinfo  {journal}
  {Nanotechnol.}\ }\textbf {\bibinfo {volume} {15}},\ \bibinfo {pages} {S472}
  (\bibinfo {year} {2004})}\BibitemShut {NoStop}%
\bibitem [{\citenamefont {Agra{\"{i}}t}\ \emph {et~al.}(2002)\citenamefont
  {Agra{\"{i}}t}, \citenamefont {Untiedt}, \citenamefont {Rubio-Bollinger},\
  and\ \citenamefont {Vieira}}]{Agrait2002}%
  \BibitemOpen
  \bibfield  {author} {\bibinfo {author} {\bibfnamefont {N.}~\bibnamefont
  {Agra{\"{i}}t}}, \bibinfo {author} {\bibfnamefont {C.}~\bibnamefont
  {Untiedt}}, \bibinfo {author} {\bibfnamefont {G.}~\bibnamefont
  {Rubio-Bollinger}}, \ and\ \bibinfo {author} {\bibfnamefont {S.}~\bibnamefont
  {Vieira}},\ }\href {\doibase 10.1103/PhysRevLett.88.216803} {\bibfield
  {journal} {\bibinfo  {journal} {Phys. Rev. Lett.}\ }\textbf {\bibinfo
  {volume} {88}},\ \bibinfo {pages} {4} (\bibinfo {year} {2002})}\BibitemShut
  {NoStop}%
\bibitem [{\citenamefont {Todorov}\ \emph {et~al.}(2001)\citenamefont
  {Todorov}, \citenamefont {Hoekstra},\ and\ \citenamefont
  {Sutton}}]{Todorov2001}%
  \BibitemOpen
  \bibfield  {author} {\bibinfo {author} {\bibfnamefont {T.~N.}\ \bibnamefont
  {Todorov}}, \bibinfo {author} {\bibfnamefont {J.}~\bibnamefont {Hoekstra}}, \
  and\ \bibinfo {author} {\bibfnamefont {A.~P.}\ \bibnamefont {Sutton}},\
  }\href {\doibase 10.1103/PhysRevLett.86.3606} {\bibfield  {journal} {\bibinfo
   {journal} {Phys. Rev. Lett.}\ }\textbf {\bibinfo {volume} {86}},\ \bibinfo
  {pages} {3606} (\bibinfo {year} {2001})}\BibitemShut {NoStop}%
\bibitem [{\citenamefont {Dundas}\ \emph {et~al.}(2009)\citenamefont {Dundas},
  \citenamefont {McEniry},\ and\ \citenamefont {Todorov}}]{Dundas2009}%
  \BibitemOpen
  \bibfield  {author} {\bibinfo {author} {\bibfnamefont {D.}~\bibnamefont
  {Dundas}}, \bibinfo {author} {\bibfnamefont {E.~J.}\ \bibnamefont {McEniry}},
  \ and\ \bibinfo {author} {\bibfnamefont {T.~N.}\ \bibnamefont {Todorov}},\
  }\href {\doibase 10.1038/nnano.2008.411} {\bibfield  {journal} {\bibinfo
  {journal} {Nat. Nanotechnol.}\ }\textbf {\bibinfo {volume} {4}},\ \bibinfo
  {pages} {99} (\bibinfo {year} {2009})}\BibitemShut {NoStop}%
\bibitem [{\citenamefont {Frederiksen}\ \emph {et~al.}(2004)\citenamefont
  {Frederiksen}, \citenamefont {Brandbyge}, \citenamefont {Lorente},\ and\
  \citenamefont {Jauho}}]{Frederiksen04}%
  \BibitemOpen
  \bibfield  {author} {\bibinfo {author} {\bibfnamefont {T.}~\bibnamefont
  {Frederiksen}}, \bibinfo {author} {\bibfnamefont {M.}~\bibnamefont
  {Brandbyge}}, \bibinfo {author} {\bibfnamefont {N.}~\bibnamefont {Lorente}},
  \ and\ \bibinfo {author} {\bibfnamefont {A.-P.}\ \bibnamefont {Jauho}},\
  }\href {\doibase 10.1103/PhysRevLett.93.256601} {\bibfield  {journal}
  {\bibinfo  {journal} {Phys. Rev. Lett.}\ }\textbf {\bibinfo {volume} {93}},\
  \bibinfo {pages} {256601} (\bibinfo {year} {2004})}\BibitemShut {NoStop}%
\bibitem [{\citenamefont {L{\"{u}}}\ \emph {et~al.}(2010)\citenamefont
  {L{\"{u}}}, \citenamefont {Brandbyge},\ and\ \citenamefont
  {Hedeg{\aa}rd}}]{Lu2010}%
  \BibitemOpen
  \bibfield  {author} {\bibinfo {author} {\bibfnamefont {J.~T.}\ \bibnamefont
  {L{\"{u}}}}, \bibinfo {author} {\bibfnamefont {M.}~\bibnamefont {Brandbyge}},
  \ and\ \bibinfo {author} {\bibfnamefont {P.}~\bibnamefont {Hedeg{\aa}rd}},\
  }\href {\doibase 10.1021/nl904233u} {\bibfield  {journal} {\bibinfo
  {journal} {Nano Lett.}\ }\textbf {\bibinfo {volume} {10}},\ \bibinfo {pages}
  {1657} (\bibinfo {year} {2010})}\BibitemShut {NoStop}%
\bibitem [{\citenamefont {L{\"{u}}}\ \emph {et~al.}(2015)\citenamefont
  {L{\"{u}}}, \citenamefont {Christensen}, \citenamefont {Wang}, \citenamefont
  {Hedeg{\aa}rd},\ and\ \citenamefont {Brandbyge}}]{Lu2015}%
  \BibitemOpen
  \bibfield  {author} {\bibinfo {author} {\bibfnamefont {J.~T.}\ \bibnamefont
  {L{\"{u}}}}, \bibinfo {author} {\bibfnamefont {R.~B.}\ \bibnamefont
  {Christensen}}, \bibinfo {author} {\bibfnamefont {J.~S.}\ \bibnamefont
  {Wang}}, \bibinfo {author} {\bibfnamefont {P.}~\bibnamefont {Hedeg{\aa}rd}},
  \ and\ \bibinfo {author} {\bibfnamefont {M.}~\bibnamefont {Brandbyge}},\
  }\href@noop {} {\bibfield  {journal} {\bibinfo  {journal} {Phys. Rev. Lett.}\
  }\textbf {\bibinfo {volume} {114}},\ \bibinfo {pages} {096801} (\bibinfo
  {year} {2015})}\BibitemShut {NoStop}%
\bibitem [{\citenamefont {Lee}\ \emph {et~al.}(2013)\citenamefont {Lee},
  \citenamefont {Kim}, \citenamefont {Jeong}, \citenamefont {Zotti},
  \citenamefont {Pauly}, \citenamefont {Cuevas},\ and\ \citenamefont
  {Reddy}}]{Lee2013}%
  \BibitemOpen
  \bibfield  {author} {\bibinfo {author} {\bibfnamefont {W.}~\bibnamefont
  {Lee}}, \bibinfo {author} {\bibfnamefont {K.}~\bibnamefont {Kim}}, \bibinfo
  {author} {\bibfnamefont {W.}~\bibnamefont {Jeong}}, \bibinfo {author}
  {\bibfnamefont {L.~A.}\ \bibnamefont {Zotti}}, \bibinfo {author}
  {\bibfnamefont {F.}~\bibnamefont {Pauly}}, \bibinfo {author} {\bibfnamefont
  {J.~C.}\ \bibnamefont {Cuevas}}, \ and\ \bibinfo {author} {\bibfnamefont
  {P.}~\bibnamefont {Reddy}},\ }\href {\doibase 10.1038/nature12183} {\bibfield
   {journal} {\bibinfo  {journal} {Nature}\ }\textbf {\bibinfo {volume}
  {498}},\ \bibinfo {pages} {209} (\bibinfo {year} {2013})}\BibitemShut
  {NoStop}%
\bibitem [{\citenamefont {Tsutsui}\ \emph {et~al.}(2018)\citenamefont
  {Tsutsui}, \citenamefont {Morikawa}, \citenamefont {Yokota},\ and\
  \citenamefont {Taniguchi}}]{Tsutsui2018}%
  \BibitemOpen
  \bibfield  {author} {\bibinfo {author} {\bibfnamefont {M.}~\bibnamefont
  {Tsutsui}}, \bibinfo {author} {\bibfnamefont {T.}~\bibnamefont {Morikawa}},
  \bibinfo {author} {\bibfnamefont {K.}~\bibnamefont {Yokota}}, \ and\ \bibinfo
  {author} {\bibfnamefont {M.}~\bibnamefont {Taniguchi}},\ }\href@noop {}
  {\bibfield  {journal} {\bibinfo  {journal} {Sci. Rep.}\ }\textbf {\bibinfo
  {volume} {8}},\ \bibinfo {pages} {7842} (\bibinfo {year} {2018})}\BibitemShut
  {NoStop}%
\bibitem [{\citenamefont {Tsutsui}\ \emph {et~al.}(2005)\citenamefont
  {Tsutsui}, \citenamefont {Taninouchi}, \citenamefont {Kurokawa},\ and\
  \citenamefont {Sakai}}]{Tsutsui2005}%
  \BibitemOpen
  \bibfield  {author} {\bibinfo {author} {\bibfnamefont {M.}~\bibnamefont
  {Tsutsui}}, \bibinfo {author} {\bibfnamefont {Y.-K.}\ \bibnamefont
  {Taninouchi}}, \bibinfo {author} {\bibfnamefont {S.}~\bibnamefont
  {Kurokawa}}, \ and\ \bibinfo {author} {\bibfnamefont {A.}~\bibnamefont
  {Sakai}},\ }\href {\doibase 10.1143/JJAP.44.5188} {\bibfield  {journal}
  {\bibinfo  {journal} {J. J. Appl. Phys.}\ }\textbf {\bibinfo {volume} {44}},\
  \bibinfo {pages} {5188} (\bibinfo {year} {2005})}\BibitemShut {NoStop}%
\bibitem [{\citenamefont {Engelund}\ \emph {et~al.}(2010)\citenamefont
  {Engelund}, \citenamefont {F{\"{u}}rst}, \citenamefont {Jauho},\ and\
  \citenamefont {Brandbyge}}]{Engelund2010}%
  \BibitemOpen
  \bibfield  {author} {\bibinfo {author} {\bibfnamefont {M.}~\bibnamefont
  {Engelund}}, \bibinfo {author} {\bibfnamefont {J.~A.}\ \bibnamefont
  {F{\"{u}}rst}}, \bibinfo {author} {\bibfnamefont {A.~P.}\ \bibnamefont
  {Jauho}}, \ and\ \bibinfo {author} {\bibfnamefont {M.}~\bibnamefont
  {Brandbyge}},\ }\href@noop {} {\bibfield  {journal} {\bibinfo  {journal}
  {Phys. Rev. Lett.}\ }\textbf {\bibinfo {volume} {104}},\ \bibinfo {pages}
  {036807} (\bibinfo {year} {2010})}\BibitemShut {NoStop}%
\bibitem [{\citenamefont {Wang}(2007)}]{Wang07}%
  \BibitemOpen
  \bibfield  {author} {\bibinfo {author} {\bibfnamefont {J.-S.}\ \bibnamefont
  {Wang}},\ }\href {\doibase 10.1103/PhysRevLett.99.160601} {\bibfield
  {journal} {\bibinfo  {journal} {Phys. Rev. Lett.}\ }\textbf {\bibinfo
  {volume} {99}},\ \bibinfo {pages} {160601} (\bibinfo {year}
  {2007})}\BibitemShut {NoStop}%
\bibitem [{\citenamefont {Bode}\ \emph {et~al.}(2011)\citenamefont {Bode},
  \citenamefont {Kusminskiy}, \citenamefont {Egger},\ and\ \citenamefont {{von
  Oppen}}}]{Bode2011}%
  \BibitemOpen
  \bibfield  {author} {\bibinfo {author} {\bibfnamefont {N.}~\bibnamefont
  {Bode}}, \bibinfo {author} {\bibfnamefont {S.~V.}\ \bibnamefont
  {Kusminskiy}}, \bibinfo {author} {\bibfnamefont {R.}~\bibnamefont {Egger}}, \
  and\ \bibinfo {author} {\bibfnamefont {F.}~\bibnamefont {{von Oppen}}},\
  }\href@noop {} {\bibfield  {journal} {\bibinfo  {journal} {Phys. Rev. Lett.}\
  }\textbf {\bibinfo {volume} {107}} (\bibinfo {year} {2011})}\BibitemShut
  {NoStop}%
\bibitem [{\citenamefont {L{\"{u}}}\ \emph {et~al.}(2012)\citenamefont
  {L{\"{u}}}, \citenamefont {Brandbyge}, \citenamefont {Hedeg{\aa}rd},
  \citenamefont {Todorov},\ and\ \citenamefont {Dundas}}]{Lu2012}%
  \BibitemOpen
  \bibfield  {author} {\bibinfo {author} {\bibfnamefont {J.~T.}\ \bibnamefont
  {L{\"{u}}}}, \bibinfo {author} {\bibfnamefont {M.}~\bibnamefont {Brandbyge}},
  \bibinfo {author} {\bibfnamefont {P.}~\bibnamefont {Hedeg{\aa}rd}}, \bibinfo
  {author} {\bibfnamefont {T.~N.}\ \bibnamefont {Todorov}}, \ and\ \bibinfo
  {author} {\bibfnamefont {D.}~\bibnamefont {Dundas}},\ }\href@noop {}
  {\bibfield  {journal} {\bibinfo  {journal} {Phys. Rev. B}\ }\textbf {\bibinfo
  {volume} {85}} (\bibinfo {year} {2012})}\BibitemShut {NoStop}%
\bibitem [{\citenamefont {Kantorovich}(2018)}]{Kantorovich2018}%
  \BibitemOpen
  \bibfield  {author} {\bibinfo {author} {\bibfnamefont {L.}~\bibnamefont
  {Kantorovich}},\ }\href {\doibase 10.1103/PhysRevB.98.014307} {\bibfield
  {journal} {\bibinfo  {journal} {Phys. Rev. B}\ }\textbf {\bibinfo {volume}
  {98}},\ \bibinfo {pages} {1} (\bibinfo {year} {2018})}\BibitemShut {NoStop}%
\bibitem [{\citenamefont {Chen}\ \emph {et~al.}(2019)\citenamefont {Chen},
  \citenamefont {Miwa},\ and\ \citenamefont {Galperin}}]{Chen2019}%
  \BibitemOpen
  \bibfield  {author} {\bibinfo {author} {\bibfnamefont {F.}~\bibnamefont
  {Chen}}, \bibinfo {author} {\bibfnamefont {K.}~\bibnamefont {Miwa}}, \ and\
  \bibinfo {author} {\bibfnamefont {M.}~\bibnamefont {Galperin}},\ }\href
  {\doibase 10.1021/acs.jpca.8b09251} {\bibfield  {journal} {\bibinfo
  {journal} {J. Phys. Chem. A}\ }\textbf {\bibinfo {volume} {123}},\ \bibinfo
  {pages} {693} (\bibinfo {year} {2019})}\BibitemShut {NoStop}%
\bibitem [{\citenamefont {L{\"{u}}}\ \emph {et~al.}(2019)\citenamefont
  {L{\"{u}}}, \citenamefont {Hu}, \citenamefont {Hedeg{\aa}rd},\ and\
  \citenamefont {Brandbyge}}]{Lu2019}%
  \BibitemOpen
  \bibfield  {author} {\bibinfo {author} {\bibfnamefont {J.~T.}\ \bibnamefont
  {L{\"{u}}}}, \bibinfo {author} {\bibfnamefont {B.~Z.}\ \bibnamefont {Hu}},
  \bibinfo {author} {\bibfnamefont {P.}~\bibnamefont {Hedeg{\aa}rd}}, \ and\
  \bibinfo {author} {\bibfnamefont {M.}~\bibnamefont {Brandbyge}},\ }\href
  {\doibase 10.1016/j.progsurf.2018.07.002} {\bibfield  {journal} {\bibinfo
  {journal} {Prog. Surf. Sci.}\ }\textbf {\bibinfo {volume} {94}},\ \bibinfo
  {pages} {21} (\bibinfo {year} {2019})}\BibitemShut {NoStop}%
\bibitem [{Note1()}]{Note1}%
  \BibitemOpen
  \bibinfo {note} {We use SIESTA\cite {Soler2002}, TranSIESTA\cite
  {Brandbyge2002} and Inelastica \cite {Frederiksen07} tool set to get the
  electronic, vibrational spectrum and their coupling. For the MD, we use the
  i-PI software\cite {ipi} (cf. SM)}\BibitemShut {NoStop}%
\bibitem [{\citenamefont {Frederiksen}\ \emph {et~al.}(2007)\citenamefont
  {Frederiksen}, \citenamefont {Paulsson}, \citenamefont {Brandbyge},\ and\
  \citenamefont {Jauho}}]{Frederiksen07}%
  \BibitemOpen
  \bibfield  {author} {\bibinfo {author} {\bibfnamefont {T.}~\bibnamefont
  {Frederiksen}}, \bibinfo {author} {\bibfnamefont {M.}~\bibnamefont
  {Paulsson}}, \bibinfo {author} {\bibfnamefont {M.}~\bibnamefont {Brandbyge}},
  \ and\ \bibinfo {author} {\bibfnamefont {A.-P.}\ \bibnamefont {Jauho}},\
  }\href {\doibase 10.1103/PhysRevB.75.205413} {\bibfield  {journal} {\bibinfo
  {journal} {Phys. Rev. B}\ }\textbf {\bibinfo {volume} {75}},\ \bibinfo
  {pages} {205413} (\bibinfo {year} {2007})}\BibitemShut {NoStop}%
\bibitem [{Note2()}]{Note2}%
  \BibitemOpen
  \bibinfo {note} {In order to clearly observe the effect we have excluded the
  change of the atomic structure and thus the bond length\cite
  {Brandbyge2003}.}\BibitemShut {Stop}%
\bibitem [{\citenamefont {Cheng}\ \emph {et~al.}(2011)\citenamefont {Cheng},
  \citenamefont {Skouta}, \citenamefont {Vazquez}, \citenamefont {Widawsky},
  \citenamefont {Schneebeli}, \citenamefont {Chen}, \citenamefont {Hybertsen},
  \citenamefont {Breslow},\ and\ \citenamefont {Venkataraman}}]{Cheng2011}%
  \BibitemOpen
  \bibfield  {author} {\bibinfo {author} {\bibfnamefont {Z.~L.}\ \bibnamefont
  {Cheng}}, \bibinfo {author} {\bibfnamefont {R.}~\bibnamefont {Skouta}},
  \bibinfo {author} {\bibfnamefont {H.}~\bibnamefont {Vazquez}}, \bibinfo
  {author} {\bibfnamefont {J.~R.}\ \bibnamefont {Widawsky}}, \bibinfo {author}
  {\bibfnamefont {S.}~\bibnamefont {Schneebeli}}, \bibinfo {author}
  {\bibfnamefont {W.}~\bibnamefont {Chen}}, \bibinfo {author} {\bibfnamefont
  {M.~S.}\ \bibnamefont {Hybertsen}}, \bibinfo {author} {\bibfnamefont
  {R.}~\bibnamefont {Breslow}}, \ and\ \bibinfo {author} {\bibfnamefont
  {L.}~\bibnamefont {Venkataraman}},\ }\href {\doibase 10.1038/nnano.2011.66}
  {\bibfield  {journal} {\bibinfo  {journal} {Nat. Nanotechnol.}\ }\textbf
  {\bibinfo {volume} {6}},\ \bibinfo {pages} {353} (\bibinfo {year}
  {2011})}\BibitemShut {NoStop}%
\bibitem [{\citenamefont {Tsutsui}\ \emph {et~al.}(2008)\citenamefont
  {Tsutsui}, \citenamefont {Taniguchi},\ and\ \citenamefont
  {Kawai}}]{Tsutsui2008}%
  \BibitemOpen
  \bibfield  {author} {\bibinfo {author} {\bibfnamefont {M.}~\bibnamefont
  {Tsutsui}}, \bibinfo {author} {\bibfnamefont {M.}~\bibnamefont {Taniguchi}},
  \ and\ \bibinfo {author} {\bibfnamefont {T.}~\bibnamefont {Kawai}},\ }\href
  {\doibase 10.1021/nl801669e} {\bibfield  {journal} {\bibinfo  {journal} {Nano
  Lett.}\ }\textbf {\bibinfo {volume} {8}},\ \bibinfo {pages} {3293} (\bibinfo
  {year} {2008})}\BibitemShut {NoStop}%
\bibitem [{\citenamefont {Tsutsui}\ \emph {et~al.}(2007)\citenamefont
  {Tsutsui}, \citenamefont {Kurokawa},\ and\ \citenamefont
  {Sakai}}]{Tsutsui2007}%
  \BibitemOpen
  \bibfield  {author} {\bibinfo {author} {\bibfnamefont {M.}~\bibnamefont
  {Tsutsui}}, \bibinfo {author} {\bibfnamefont {S.}~\bibnamefont {Kurokawa}}, \
  and\ \bibinfo {author} {\bibfnamefont {A.}~\bibnamefont {Sakai}},\ }\href
  {\doibase 10.1063/1.2719682} {\bibfield  {journal} {\bibinfo  {journal}
  {Appl. Phys. Lett.}\ }\textbf {\bibinfo {volume} {90}},\ \bibinfo {pages}
  {133121} (\bibinfo {year} {2007})}\BibitemShut {NoStop}%
\bibitem [{\citenamefont {Erpenbeck}\ \emph {et~al.}(2018)\citenamefont
  {Erpenbeck}, \citenamefont {Schinabeck}, \citenamefont {Peskin},\ and\
  \citenamefont {Thoss}}]{Erp2018}%
  \BibitemOpen
  \bibfield  {author} {\bibinfo {author} {\bibfnamefont {A.}~\bibnamefont
  {Erpenbeck}}, \bibinfo {author} {\bibfnamefont {C.}~\bibnamefont
  {Schinabeck}}, \bibinfo {author} {\bibfnamefont {U.}~\bibnamefont {Peskin}},
  \ and\ \bibinfo {author} {\bibfnamefont {M.}~\bibnamefont {Thoss}},\ }\href
  {\doibase 10.1103/PhysRevB.97.235452} {\bibfield  {journal} {\bibinfo
  {journal} {Phys. Rev. B}\ }\textbf {\bibinfo {volume} {97}},\ \bibinfo
  {pages} {235452} (\bibinfo {year} {2018})}\BibitemShut {NoStop}%
\bibitem [{\citenamefont {Pobelov}\ \emph {et~al.}(2017)\citenamefont
  {Pobelov}, \citenamefont {Lauritzen}, \citenamefont {Yoshida}, \citenamefont
  {Jensen}, \citenamefont {Meszaros}, \citenamefont {Jacobsen}, \citenamefont
  {Strange}, \citenamefont {Wandlowski},\ and\ \citenamefont
  {Solomon}}]{solomon2017}%
  \BibitemOpen
  \bibfield  {author} {\bibinfo {author} {\bibfnamefont {I.~V.}\ \bibnamefont
  {Pobelov}}, \bibinfo {author} {\bibfnamefont {K.~P.}\ \bibnamefont
  {Lauritzen}}, \bibinfo {author} {\bibfnamefont {K.}~\bibnamefont {Yoshida}},
  \bibinfo {author} {\bibfnamefont {A.}~\bibnamefont {Jensen}}, \bibinfo
  {author} {\bibfnamefont {G.}~\bibnamefont {Meszaros}}, \bibinfo {author}
  {\bibfnamefont {K.~W.}\ \bibnamefont {Jacobsen}}, \bibinfo {author}
  {\bibfnamefont {M.}~\bibnamefont {Strange}}, \bibinfo {author} {\bibfnamefont
  {T.}~\bibnamefont {Wandlowski}}, \ and\ \bibinfo {author} {\bibfnamefont
  {G.~C.}\ \bibnamefont {Solomon}},\ }\href {\doibase 10.1038/ncomms15931}
  {\bibfield  {journal} {\bibinfo  {journal} {Nat. Commun.}\ }\textbf {\bibinfo
  {volume} {8}},\ \bibinfo {pages} {15931} (\bibinfo {year}
  {2017})}\BibitemShut {NoStop}%
\bibitem [{\citenamefont {Soler}\ \emph {et~al.}(2002)\citenamefont {Soler},
  \citenamefont {Artacho}, \citenamefont {Gale}, \citenamefont {Garc{\'{i}}a},
  \citenamefont {Junquera}, \citenamefont {Ordej{\'{o}}n},\ and\ \citenamefont
  {S{\'{a}}nchez-Portal}}]{Soler2002}%
  \BibitemOpen
  \bibfield  {author} {\bibinfo {author} {\bibfnamefont {J.~M.}\ \bibnamefont
  {Soler}}, \bibinfo {author} {\bibfnamefont {E.}~\bibnamefont {Artacho}},
  \bibinfo {author} {\bibfnamefont {J.~D.}\ \bibnamefont {Gale}}, \bibinfo
  {author} {\bibfnamefont {A.}~\bibnamefont {Garc{\'{i}}a}}, \bibinfo {author}
  {\bibfnamefont {J.}~\bibnamefont {Junquera}}, \bibinfo {author}
  {\bibfnamefont {P.}~\bibnamefont {Ordej{\'{o}}n}}, \ and\ \bibinfo {author}
  {\bibfnamefont {D.}~\bibnamefont {S{\'{a}}nchez-Portal}},\ }\href@noop {}
  {\bibfield  {journal} {\bibinfo  {journal} {J. Phys. Condens. Matt.}\
  }\textbf {\bibinfo {volume} {14}},\ \bibinfo {pages} {2745} (\bibinfo {year}
  {2002})}\BibitemShut {NoStop}%
\bibitem [{\citenamefont {Brandbyge}\ \emph {et~al.}(2002)\citenamefont
  {Brandbyge}, \citenamefont {Mozos}, \citenamefont {Ordej{\'{o}}n},
  \citenamefont {Taylor},\ and\ \citenamefont {Stokbro}}]{Brandbyge2002}%
  \BibitemOpen
  \bibfield  {author} {\bibinfo {author} {\bibfnamefont {M.}~\bibnamefont
  {Brandbyge}}, \bibinfo {author} {\bibfnamefont {J.~L.}\ \bibnamefont
  {Mozos}}, \bibinfo {author} {\bibfnamefont {P.}~\bibnamefont
  {Ordej{\'{o}}n}}, \bibinfo {author} {\bibfnamefont {J.}~\bibnamefont
  {Taylor}}, \ and\ \bibinfo {author} {\bibfnamefont {K.}~\bibnamefont
  {Stokbro}},\ }\href@noop {} {\bibfield  {journal} {\bibinfo  {journal} {Phys.
  Rev. B}\ }\textbf {\bibinfo {volume} {65}},\ \bibinfo {pages} {1654011}
  (\bibinfo {year} {2002})}\BibitemShut {NoStop}%
\bibitem [{\citenamefont {Kapil}\ \emph {et~al.}(2019)\citenamefont {Kapil},
  \citenamefont {Rossi}, \citenamefont {Marsalek}, \citenamefont {Petraglia},
  \citenamefont {Litman}, \citenamefont {Spura}, \citenamefont {Cheng},
  \citenamefont {Cuzzocrea}, \citenamefont {Mei\ss~ner}, \citenamefont
  {Wilkins}, \citenamefont {Helfrecht}, \citenamefont {Juda}, \citenamefont
  {Bienvenue}, \citenamefont {Fang}, \citenamefont {Kessler}, \citenamefont
  {Poltavsky}, \citenamefont {Vandenbrande}, \citenamefont {Wieme},
  \citenamefont {Corminboeuf}, \citenamefont {K\"uhne}, \citenamefont
  {Manolopoulos}, \citenamefont {Markland}, \citenamefont {Richardson},
  \citenamefont {Tkatchenko}, \citenamefont {Tribello}, \citenamefont
  {Van~Speybroeck},\ and\ \citenamefont {Ceriotti}}]{ipi}%
  \BibitemOpen
  \bibfield  {author} {\bibinfo {author} {\bibfnamefont {V.}~\bibnamefont
  {Kapil}}, \bibinfo {author} {\bibfnamefont {M.}~\bibnamefont {Rossi}},
  \bibinfo {author} {\bibfnamefont {O.}~\bibnamefont {Marsalek}}, \bibinfo
  {author} {\bibfnamefont {R.}~\bibnamefont {Petraglia}}, \bibinfo {author}
  {\bibfnamefont {Y.}~\bibnamefont {Litman}}, \bibinfo {author} {\bibfnamefont
  {T.}~\bibnamefont {Spura}}, \bibinfo {author} {\bibfnamefont
  {B.}~\bibnamefont {Cheng}}, \bibinfo {author} {\bibfnamefont
  {A.}~\bibnamefont {Cuzzocrea}}, \bibinfo {author} {\bibfnamefont {R.~H.}\
  \bibnamefont {Mei\ss~ner}}, \bibinfo {author} {\bibfnamefont {D.~M.}\
  \bibnamefont {Wilkins}}, \bibinfo {author} {\bibfnamefont {B.~A.}\
  \bibnamefont {Helfrecht}}, \bibinfo {author} {\bibfnamefont {P.}~\bibnamefont
  {Juda}}, \bibinfo {author} {\bibfnamefont {S.~P.}\ \bibnamefont {Bienvenue}},
  \bibinfo {author} {\bibfnamefont {W.}~\bibnamefont {Fang}}, \bibinfo {author}
  {\bibfnamefont {J.}~\bibnamefont {Kessler}}, \bibinfo {author} {\bibfnamefont
  {I.}~\bibnamefont {Poltavsky}}, \bibinfo {author} {\bibfnamefont
  {S.}~\bibnamefont {Vandenbrande}}, \bibinfo {author} {\bibfnamefont
  {J.}~\bibnamefont {Wieme}}, \bibinfo {author} {\bibfnamefont
  {C.}~\bibnamefont {Corminboeuf}}, \bibinfo {author} {\bibfnamefont {T.~D.}\
  \bibnamefont {K\"uhne}}, \bibinfo {author} {\bibfnamefont {D.~E.}\
  \bibnamefont {Manolopoulos}}, \bibinfo {author} {\bibfnamefont {T.~E.}\
  \bibnamefont {Markland}}, \bibinfo {author} {\bibfnamefont {J.~O.}\
  \bibnamefont {Richardson}}, \bibinfo {author} {\bibfnamefont
  {A.}~\bibnamefont {Tkatchenko}}, \bibinfo {author} {\bibfnamefont {G.~A.}\
  \bibnamefont {Tribello}}, \bibinfo {author} {\bibfnamefont {V.}~\bibnamefont
  {Van~Speybroeck}}, \ and\ \bibinfo {author} {\bibfnamefont {M.}~\bibnamefont
  {Ceriotti}},\ }\href {\doibase https://doi.org/10.1016/j.cpc.2018.09.020}
  {\bibfield  {journal} {\bibinfo  {journal} {Comp. Phys. Comm.}\ }\textbf
  {\bibinfo {volume} {236}},\ \bibinfo {pages} {214} (\bibinfo {year}
  {2019})}\BibitemShut {NoStop}%
\bibitem [{\citenamefont {Brandbyge}\ \emph {et~al.}(2003)\citenamefont
  {Brandbyge}, \citenamefont {Stokbro}, \citenamefont {Taylor}, \citenamefont
  {Mozos},\ and\ \citenamefont {Ordej{\'{o}}n}}]{Brandbyge2003}%
  \BibitemOpen
  \bibfield  {author} {\bibinfo {author} {\bibfnamefont {M.}~\bibnamefont
  {Brandbyge}}, \bibinfo {author} {\bibfnamefont {K.}~\bibnamefont {Stokbro}},
  \bibinfo {author} {\bibfnamefont {J.}~\bibnamefont {Taylor}}, \bibinfo
  {author} {\bibfnamefont {J.~L.}\ \bibnamefont {Mozos}}, \ and\ \bibinfo
  {author} {\bibfnamefont {P.}~\bibnamefont {Ordej{\'{o}}n}},\ }\href@noop {}
  {\bibfield  {journal} {\bibinfo  {journal} {Phys. Rev. B}\ }\textbf {\bibinfo
  {volume} {67}},\ \bibinfo {pages} {193104} (\bibinfo {year}
  {2003})}\BibitemShut {NoStop}%
\bibitem [{\citenamefont {Engelund}\ \emph {et~al.}(2009)\citenamefont
  {Engelund}, \citenamefont {Brandbyge},\ and\ \citenamefont
  {Jauho}}]{Engelund2009}%
  \BibitemOpen
  \bibfield  {author} {\bibinfo {author} {\bibfnamefont {M.}~\bibnamefont
  {Engelund}}, \bibinfo {author} {\bibfnamefont {M.}~\bibnamefont {Brandbyge}},
  \ and\ \bibinfo {author} {\bibfnamefont {A.~P.}\ \bibnamefont {Jauho}},\
  }\href {https://link.aps.org/doi/10.1103/PhysRevB.80.045427} {\bibfield
  {journal} {\bibinfo  {journal} {Phys. Rev. B}\ }\textbf {\bibinfo {volume}
  {80}},\ \bibinfo {pages} {045427} (\bibinfo {year} {2009})}\BibitemShut
  {NoStop}%
\end{thebibliography}%

\clearpage
\onecolumngrid

\numberwithin{equation}{section}
\setcounter{equation}{0}
\setcounter{figure}{0}
\renewcommand\theequation{S\arabic{equation}}
\renewcommand\thefigure{S\arabic{figure}}

\begin{center}\textbf{Supplementary Materials for ``Ab initio current-induced molecular dynamics''}\end{center}
\subsection{Velocity correlation and power spectrum}
To study the energy distribution of the system at steady state, we define the velocity correlation function as
\begin{equation}
C_{v_iv_j}(t) = \langle v_i(t) v_j(0)\rangle = \frac{1}{T}\int_{0}^T dt' v_i(t'+t) v_j(t').
\label{eq:cvv}
\end{equation}
Here, $T$ is the total time of one MD trajectory, and $v(t)=\dot{Q}(t)$ the mass-normalized velocity. Using the Wiener-Khinchin theorem, the power spectrum is then given by the Fourier transform of the velocity correlation function
\begin{equation}
\frac{1}{T}\langle |v(\omega)|^2 \rangle = C_{vv}(\omega) = \sum_i C_{v_iv_i}(\omega). 
\end{equation}
with
\begin{equation}
{C}_{v_iv_i}(\omega) = \int dt\ {C}_{v_iv_i}(t) e^{i\omega t}.
\end{equation}
The cumulative integration of the power spectrum
\begin{equation}
	E_K(\omega) = \int_0^{\omega} \frac{d\omega'}{2\pi} {C}_{vv}(\omega'),
	\label{eq:ek}
\end{equation}
gives the kinetic energy contributed by modes with frequency $\le \omega$. Thus, $E_K(+\infty)$ gives the total kinetic energy. 
To characterize the real space distribution, 
we define the velocity of the bond between atoms $n$ and $k$ as
\begin{equation}
\bar v_{\alpha}^{nk} =v_{\alpha}^{n}-v_{\alpha}^{k} ,    
\end{equation}
where $\alpha=x, y, z$.
The kinetic energy of the bond is then 
\begin{equation}
    E^{b,nk} = \int_0^{+\infty} \frac{d\omega}{2\pi}\ \sum_{\alpha} \tilde{C}_{\bar{v}_\alpha^{nk}\bar{v}_\alpha^{nk}}(\omega).
    \label{eq:bond}
\end{equation}

\subsection{Validation of the method}
\begin{figure}[h!]
\centering
\includegraphics[width=0.5\textwidth]{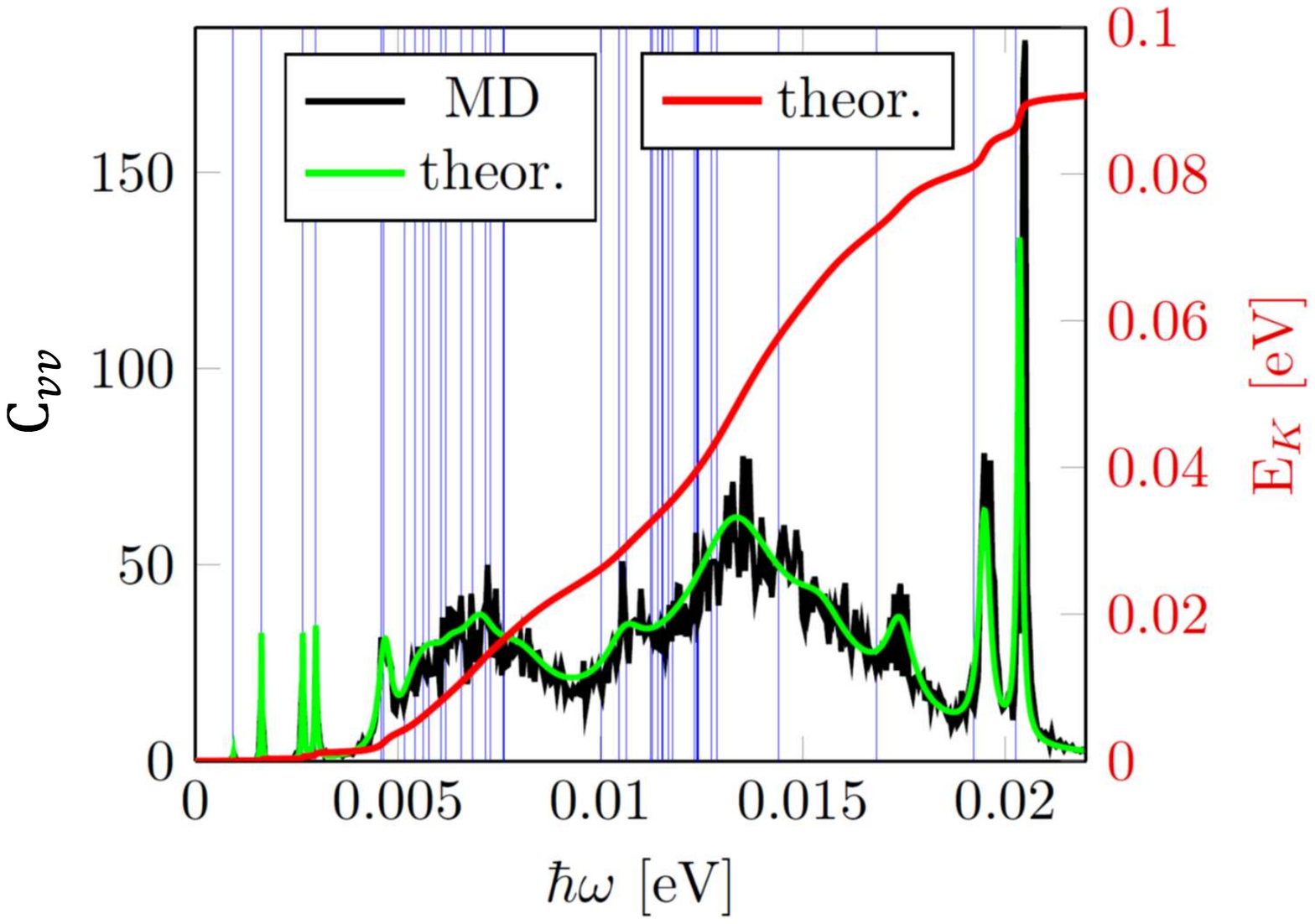} 
\caption{The power spectrum  and its cumulative integral for the 5-atom gold chain within the harmonic approximation at 0 V. The green line is the exact theoretical result from solving the SCGLE, while the noisy black lines are results of the MD simulation. They agree well with each other. The blue vertical lines mark the energy of different vibrational modes for the isolated chain. The red line is the cumulative integral of the power spectrum.}
\label{fig:au}
\end{figure}

To validate our method, we show in Fig.~\ref{fig:au} the comparison between MD and theoretical results at zero bias $V=0$ within the harmonic approximation where the SCGLE can be solved exactly. We find good agreement between the MD (black) and theoretical (green) results.  

\subsection{Effective temperature}

\begin{figure}[h!]
	\includegraphics[width=\textwidth]{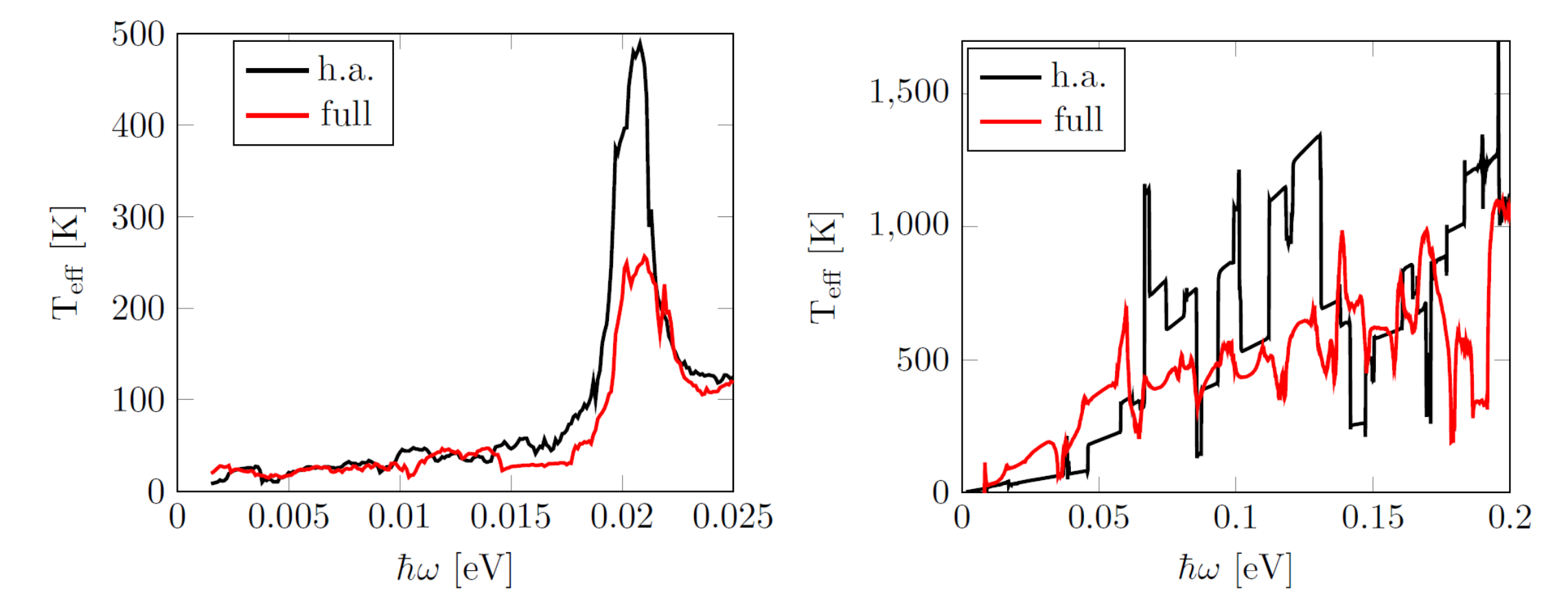}
	\caption{Effective temperature over frequencies for the Au chain (left) and the Benzene molecule (right) for $T_0 = 4$ K , $V=1.0$ V.}
	\label{fig:teff}
\end{figure}

We estimate the effective temperature  $T_{\text{eff}}$ from
\begin{align*}
 \frac{C_{vv}(\omega, V) }{C_{vv}(\omega, 0)} = \frac{2 n_B (\omega, T_{\text{eff}}) +1 }{ 2 n_B (\omega, T_0)+1}
\end{align*}
with
\begin{align*}
 n_B(\omega, T_{\text{eff}}) = \frac{1}{ e^{\frac{\hbar \omega}{k_B T_{\text{eff}}} }-1}
\end{align*}
the Bose-Einstein distribution, $k_B$ Boltzmann's constant, $\omega$ the phonon frequency,  $T_0$ the temperature of the MD simulation and $V$ the applied bias voltage. The results for Au chain and benzene molecule are shown in Fig.~\ref{fig:teff} (a) and (b), respectively.

\subsection{Computational details}
The electronic structure of the relaxed structures is calculated using \siesta\cite{Soler2002} and \tsiesta\cite{Brandbyge2002}. The vibrational spectrum and the electron-vibration coupling of the system are calculated using Inelastica\cite{Frederiksen07}. The electrode phonon self-energy is calculated using the method described in Ref.~\onlinecite{Engelund2009}. From these, all parameters of the Langevin equation can be calculated. The details of the theory can be found in Ref.~\onlinecite{Lu2012}.

In the harmonic approximation, the dynamical matrix of the central system is used to calculate the potential force $\bm{F}(\bm{Q}(t))$ in Eq. (\ref{eq:lang}), while for the full AIMD the anharmonic force is obtained from the \abinitio code \siesta. 
Specifically, to solve Eq. (\ref{eq:lang}), we are using \siesta\ in combination with the i-PI code\cite{ipi} , which is a python-based force engine for atomistic simulations using interatomic forces and potential energies calculated by an external driver code.
This means that i-PI deals with the propagation of the nuclear motion, while the external driver \siesta\ is used to compute the energy and forces. 
Communication between the client and i-PI runs through a socket interface.
\siesta\ receives the atomic coordinates from the python interface and returns forces $\bm{F}(\bm{Q}(t))$ that are used to integrate the Langevin equation, Eq. (\ref{eq:lang}),
which we have implemented in the integrator class of i-PI. 
For the MD in the harmonic approximation, our Langevin-i-PI code communicates with an external driver that calculates the forces from the dynamical matrix.

\begin{figure}[h!]
	\centering
	\includegraphics[width=0.45\textwidth]{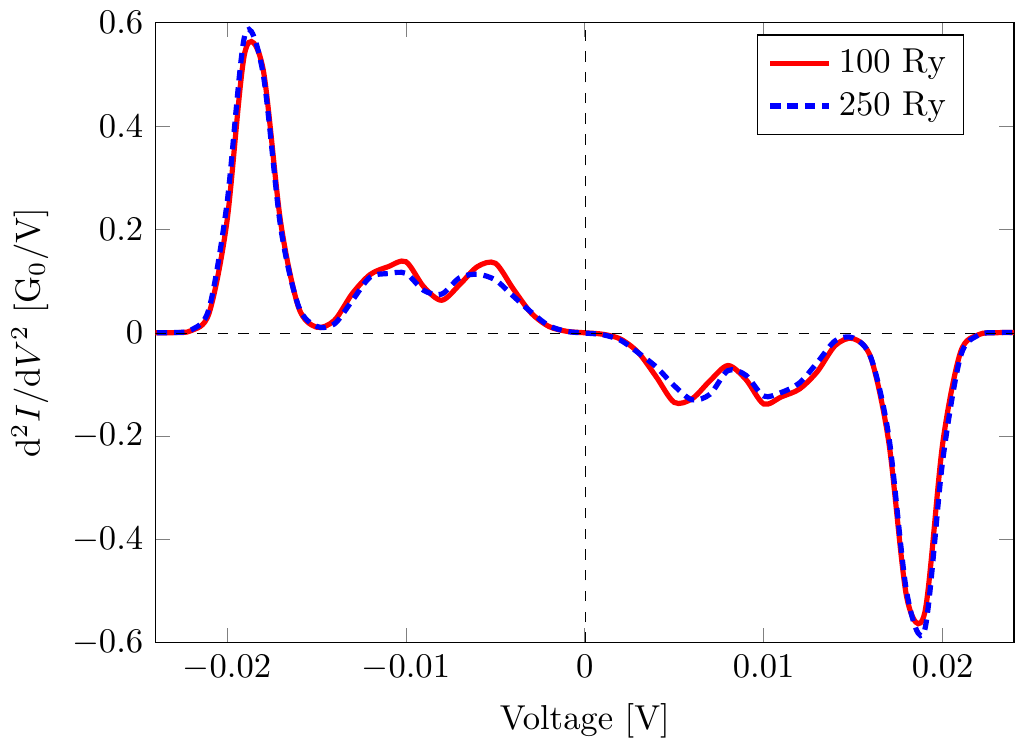}
	\caption{IETS spectrum for Au chain using different mesh cut offs. They agree reasonably well with each other. To save the computational time, we have used a mesh cut off of 100 Ry in the MD simulation.}
	\label{fig:au-iets}
\end{figure}

In the \siesta\ calculations, a single zeta polarized basis set is used for Au, and double zeta polarized basis is used for C, H. A mesh cut off of 100 Ry is used to accelerate the MD simulation. 
A comparison of the IETS of the Au chain using a cut off of 250 Ry to 100 Ry is depicted in Fig. \ref{fig:au-iets}, showing only slight influence of the cut off. 
A time step of 10 fs (1 fs) and total of 8192 (131072) MD steps are used for Au chain (benzene junction) for each independent MD run. The results shown in the main text are average over 10 (4) independent MD runs. 

\end{document}